\documentclass[twocolumn,prl,aps,superscriptaddress]{revtex4}

\usepackage{amssymb}
\usepackage{amsfonts}
\usepackage{amsmath}
\usepackage{bm}
\usepackage{CJK}
\usepackage{graphicx}        
\usepackage[usenames,dvipsnames]{xcolor}
\usepackage{siunitx}
\usepackage{mathtools}
\usepackage[section,numberedsection=autolabel]{glossaries}
\usepackage[linkcolor=violet,citecolor=blue,colorlinks=true,urlcolor=purple]{hyperref}

\newcommand{\ssg}{1s$\sigma_g$}
\newcommand{\psu}{2p$\sigma_u$}

\def\cep{\textrm{CEP}}
\def\CEP{CEP}

\def\ilphase{\phi_i^L}
\def\imphase{\phi_i^M}

\def\protonmom{\vec{p}_{\textrm{H}^+}}

\def\protonmomy{{p}_{\textrm{H}^+,y}}
\def\protonmomz{{p}_{\textrm{H}^+,z}}

\def\molfaniso{\mathcal{A}_M}
\def\aniso{\mathcal{A}}
\def\labfaniso{\mathcal{A}_L}
\def\aniso{\mathcal{A}}
\def\momdis{\mathcal{M}}

\makeglossaries
\newacronym{cep}{CEP}{carrier-envelope phase}
\newacronym{req}{$R_{eq}$}{equilibrium internuclear distance}

\newcommand{\SM}{{Suppl. Mat. \cite{SMKangaparambil2020}}}

\renewcommand{\vec}{\textbf}

\begin{document}
\begin{CJK*}{UTF8}{gbsn}

\title{Generalized phase-sensitivity of directional bond-breaking in laser-molecule interaction}


\author{Sarayoo\,Kangaparambil}
\email[]{sarayoo.kangaparambil@tuwien.ac.at}
\affiliation{Photonics Institute, Technische Universit\"at Wien, 1040 Vienna, Austria}

\author{V\'{a}clav\,Hanus}
\affiliation{Photonics Institute, Technische Universit\"at Wien, 1040 Vienna, Austria}

\author{Martin\,Dorner-Kirchner}
\affiliation{Photonics Institute, Technische Universit\"at Wien, 1040 Vienna, Austria}

\author{Peilun He}
\affiliation{Key Laboratory for Laser Plasmas and School of Physics and Astronomy, Shanghai Jiao Tong University, Shanghai 200240, China}

\author{Seyedreza\,Larimian}
\affiliation{Photonics Institute, Technische Universit\"at Wien, 1040 Vienna, Austria}


\author{Gerhard\,Paulus}
\affiliation{Institute of Optics and Quantum Electronics, Friedrich Schiller University Jena, 07743 Jena, Germany}

\author{Andrius\,Baltu\v{s}ka}
\affiliation{Photonics Institute, Technische Universit\"at Wien, 1040 Vienna, Austria}

\author{Xinhua\,Xie\,\mbox{(谢新华)}}
\affiliation{Photonics Institute, Technische Universit\"at Wien, 1040 Vienna, Austria}

\author{Kaoru\,Yamanouchi}
\affiliation{Department of Chemistry, School of Science, The University of Tokyo, 7-3-1 Hongo, Bunkyo-ku, Tokyo 113-0033, Japan}

\author{Feng\,He}
\affiliation{Key Laboratory for Laser Plasmas and School of Physics and Astronomy, Shanghai Jiao Tong University, Shanghai 200240, China}


\author{Erik\,L\"otstedt}
\affiliation{Department of Chemistry, School of Science, The University of Tokyo, 7-3-1 Hongo, Bunkyo-ku, Tokyo 113-0033, Japan}

\author{Markus\,Kitzler-Zeiler}
\email[]{markus.kitzler-zeiler@tuwien.ac.at}
\affiliation{Photonics Institute, Technische Universit\"at Wien, 1040 Vienna, Austria}

\begin{abstract}
We establish a generalized picture of the phase-sensitivity of laser-induced directional bond-breaking using the H$_2$ molecule as the example. We show that the well-known proton ejection anisotropy measured with few-cycle pulses arises as an amplitude-modulation of an intrinsic anisotropy that is sensitive to the laser phase at the ionization time and determined by the molecule's electronic structure. Our work furthermore reveals a strong electron-proton correlation that may open up a new approach to experimentally accessing the laser-sub-cycle intramolecular electron dynamics also in large molecules.
\end{abstract}


\maketitle
\end{CJK*}

Bond-breaking in molecules with intense laser pulses is a well-established field of science 
 \cite{Sheehy2001, Posthumus2004, Kling2013, Li2017, Alnaser2017, Ibrahim2018}. 
An important step in this field was the demonstration of anisotropic proton ejection during dissociative laser ionization of H$_2$ using the \gls{cep} of intense few-cycle pulses \cite{Kling2006, Kremer2009, Fischer2010, Znakovskaya2012, Xu2013,  Fischer2013, Kling2013a, Li2016} or the relative phase between two colors of multi-cycle pulses \cite{Ray2009, Wanie2016, Xie2017b}. 
\gls{cep} is, however, not the only phase that can determine the directionality of a bond-breaking reaction. 
A pronounced anisotropy in proton ejection from H$_2$ can also be observed with single-color many-cycle fields as a function of the laser phase at the instant of ionization \cite{Wu2013}. 
%
%
%
%
However, the relation of these two phases for determining the ejection direction of the proton is unclear. Moreover, the fact that the instant of electron emission within a laser cycle can determine the directionality of bond-breaking, questions the role of the \gls{cep}-dependent laser field shape after the ionization step in the formation of the proton anisotropy.


In this Letter, we establish a unified picture that connects the roles of the ionization phase and the \CEP{} in determining the directionality of proton ejection in laser-dissociation of H$_2$. 
We show that the ejection anisotropy due to the ionization phase is an intrinsic property of the molecule that can only be observed in the \textit{molecular} frame. In contrast, the \CEP{} acts in the \textit{lab} frame. 
Our experiments and simulations prove that the connection between the action of the two phases is provided by the \CEP{}-modulation of the instantaneous laser field strength at the time of population transfer between electronic states of H$_2^+$. 
%
%
Moreover, our experiments reveal a remarkably strong correlation between the outgoing electron and the ejected proton that is due to the intramolecular electron dynamics during dissociation. This finding may open up a new approach to investigating laser-sub-cycle electronic dynamics also in larger molecules.



\begin{figure*}[tb]
 \centering
  \includegraphics[width=0.99\textwidth]{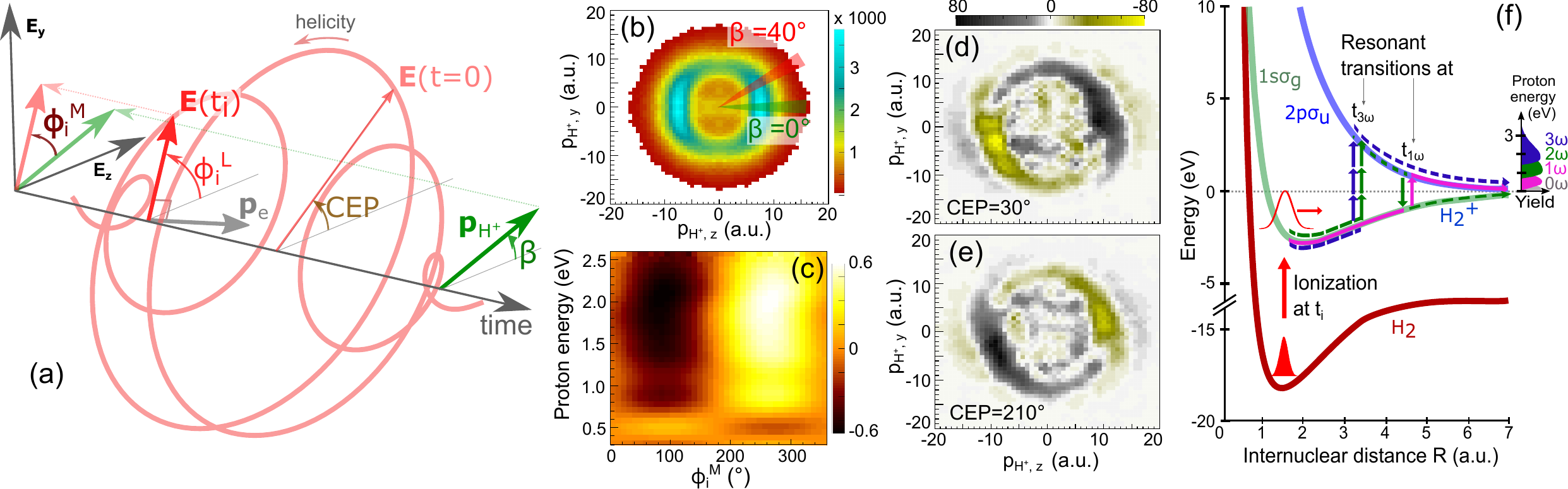}
    \caption{(a) Evolution of laser electric field vector $\vec{E}(t)$ over time, $\vec{E}(t_i)$ at ionization time $t_i$ (derived from the electron momentum vector $\vec{p}_e$), an example proton vector $\protonmom$ aligned under $\beta$ to the $z$ axis of the lab frame, the molecular frame ionization phase $\imphase$ and the lab frame ionization phase $\ilphase=\omega t_i + \cep$. 
    (b) Measured proton momentum distribution in the lab frame, integrated over \CEP{}, and the two $\beta$-ranges of proton ejection selected for Figs.~\ref{fig2}(a,b). 
    (c) Molecular frame proton anisotropy $\molfaniso$ as a function of $\imphase$ and proton kinetic energy $E_k$, integrated over the \CEP{}.
    (d,e) Anisotropy of the proton momentum distribution, $\Delta \momdis$, for $\cep=30(\pm10)^\circ$ (d) and $\cep=210(\pm10)^\circ$ (e), as detailed in the text.
    (f) Potential energy curves of H$_2$ and H$_2^+$, dissociation  pathways $n\omega$ for $n=\{1,2,3\}$ and schematic proton distributions for $n=\{0,1,2,3\}$.
    }
  \label{fig1}
\end{figure*}



The key to these achievements was the combination of elliptically polarized few-cycle laser pulses with known \gls{cep} in combination with coincidence detection of protons and electrons in our experiments \cite{Schoffler2016, Hanus2019, Hanus2020}.
In elliptically polarized light the ionization time, $t_i$, is mapped onto the emission angle of the photoelectron by the laser field via $\vec{p}_{e} = -\vec{A}(t_i)$ \cite{Faisal1973, Reiss1980}. Atomic units are used throughout. 
The laser vector potential $\vec{A}(t)$ is connected to the laser electric field by $\vec{A}(t)=-\int_{-\infty}^{t}\vec{E}(t')dt'$. 
Thus, by measuring in coincidence the momenta of the emitted electron, $\vec{p}_e$, and the proton ejected upon dissociation, $\protonmom$, the electric field vector at the time of ionization, $\vec{E}(t_i)$, and the angle of proton ejection in the lab frame, $\beta$, can be retrieved \citep{Maharjan2005, Wu2013, Hanus2019}, see Fig.~\ref{fig1}(a) for a visualization. 
From $\beta$ and $\vec{E}(t_i)$ the ionization phase in the molecular frame, $\imphase$, can be unambiguously derived (see sketch). Such retrieval is not straightforwardly possible for  linearly polarized pulses.


%
%

In our experiments, the laser field in the lab frame is $\vec{E}(t) =   f(t) 	[ 	\hat{E}_z \cos(\omega t + \cep)   \vec{e}_z + 	\hat{E}_y \sin(\omega t + \cep) \vec{e}_y  ]$, where the pulse envelope $f(t)$ (peak value 1) had a duration of 4.5\,fs (FWHM in intensity), the ellipticity was $\varepsilon = \hat{E}_y/\hat{E}_z = 0.85$ and the angular frequency $\omega$ was given by the spectral center wavelength $750$\,nm. The peak intensity was 0.8\,PW/cm$^{2}$  (measured \textit{in situ} \cite{Smeenk2011}).
The pulses were focused inside a reaction microscope (background pressure $< 10^{-10}$\,mbar) onto a supersonic beam of randomly oriented H$_2$ molecules. 
Duration and CEPs of the pulses were measured with a stereo electron spectrometer in phase-tagging mode \citep{Sayler2011a, Rathje2012}.
The value of $\cep=0$ was calibrated to the maximum of the H$_2^+$ yield into $p_z>0$. 
Protons and electrons created by the laser-molecule interaction were guided to two position and time sensitive detectors by weak electric (20\,V/cm) and magnetic fields (9\,G) for measuring in coincidence their momenta right after the laser pulse. In the offline data analysis the dissociation channel was selected. Further experimental details 
can be found in Refs.~\cite{Hanus2019, Hanus2020, Schoffler2016}.

The proton momentum distribution in the laser polarization plane, 
$\momdis(\protonmomz, \protonmomy)$,  
integrated over the third momentum component ${p}_{\textrm{H}^+, x}$ and all values of the \gls{cep}, is shown in Fig.~\ref{fig1}(b). 
Because the rotational motion of H$_2^+$ is slow as compared to the dissociation process, the proton momentum $\protonmom$ encodes the instantaneous orientation of the $\textrm{H}_2$ molecule before laser interaction \cite{Hanus2019, Hanus2020}.
As the molecules are isotropically oriented in the jet, the \gls{cep}-integrated momentum distribution reflects the laser intensity distribution of the elliptically polarized pulses and does not show any anisotropy along the radial direction with respect to the origin at any angle $\beta$. 

To investigate the influence of the \gls{cep} on the directionality of the bond-breaking process, we calculated the difference between the momentum distributions measured for a given \gls{cep}, $\momdis(\cep)$, and the \gls{cep}-integrated proton momentum distribution, $\tilde{\momdis}$. The resulting difference momentum distributions $\Delta \momdis = \momdis(\cep) - \tilde{\momdis}$ for $\cep = 30^\circ$ and $\cep = 210^\circ$ are plotted in Figs.~\ref{fig1}(d,e).
The $\Delta \momdis$ distributions show a pronounced $\beta$-dependent antisymmetry about the origin along the radial direction for certain ranges of $|\protonmom|$ that flips if the \CEP{} is changed by 180$^\circ$ (from $30^\circ$ to $210^\circ$).
These  $\Delta \momdis$ distributions constitute, to our knowledge, the first demonstration of a \CEP{}-dependent two-dimensional (2D) proton anisotropy in H$_2^+$ dissociation.  Thus far, such 2D-control was only demonstrated by the relative phase of 2D two-color laser fields  \cite{Gong2014c, Lin2016}.


Our goal is to relate the CEP-induced anistropy of Figs.~\ref{fig1}(d,e) to that due to the ionization phase in the molecular frame, $\imphase$, introduced by Ref.~\cite{Wu2013}. This anisotropy measured in our experiments is shown in Fig.~\ref{fig1}(c). 
For this plot, we retrieved for each proton in the momentum distribution Fig.~\ref{fig1}(b) the value  of $\beta$ from $\protonmom$, and from the electron momentum vector $\vec{p}_e$ we retrieved the lab frame  angle $\ilphase$ of the electric field vector $\vec{E}(t_i)$ at ionization  time. From $\ilphase$ and $\beta$ we calculated for each detected proton the angle  $\imphase = \ilphase - \beta$, which is the ionization phase with respect to $\protonmom$  (modulo $2\pi$) [cf. Fig.~\ref{fig1}(a)].
This allows to obtain the normalized proton anisotropy in the molecular frame, $\molfaniso$. For direct comparability we used the same definition for $\molfaniso$ as Wu~et~al. \cite{Wu2013}, namely $\molfaniso(\imphase,E_k,\beta) = \frac{N(\imphase,E_k, \beta)-N(\imphase+180^\circ,E_k,\beta)}{N(\imphase,E_k,\beta)+N(\imphase+180^\circ,E_k,\beta)}$, with $N(\imphase,E_k,\beta)$  the number of protons ejected along angle $\beta$ for given values of $\imphase$ and $E_k$. This number is compared to the number of protons ejected along the same angle $\beta$ for the opposite ionization phase $\imphase \rightarrow \imphase+180^\circ$. 
Because $\imphase$ is defined in the molecular frame, the numbers $N$ can be integrated over $\beta$. 
The resulting anisotropy, integrated over all values of the \gls{cep}, is shown in Fig.~\ref{fig1}(c).

Both anisotropies, $\Delta \momdis$ in Figs.~\ref{fig1}(d,e) and $\molfaniso$ in Fig.~\ref{fig1}(c), are the results of interferences of nuclear wave packets dissociating along different pathways $n \omega$ on the gerade \ssg{} and ungerade \psu{} electronic states that are associated with the number of effectively absorbed photons, $n=\{0,1,2,\dots\}$ \cite{Roudnev2007, Hua2009}, cf. sketch in Fig.~\ref{fig1}(f).
Electronic excitation by electron recollisions, relevant in other works \cite{Kling2006, Znakovskaya2012, Li2014d}, is suppressed in elliptical light.
Superposition of the wave packets with a relative phase $\Delta \varphi$ acquired during pathways associated with odd and even $n$ 
dictates the localization of the remaining electron in the dissociating H$_2^+$ molecule and therefore the anisotropy of proton ejection, $\aniso$, with an $n$-dependent kinetic energy $E_k=|\protonmom|^2/(2 m_p)$, with $m_p$ the proton mass, according to $\aniso \propto \sin(\Delta \varphi)$. See \SM{} for an extended description.

\begin{figure}[tb]
 \centering
  \includegraphics[width=0.99\columnwidth]{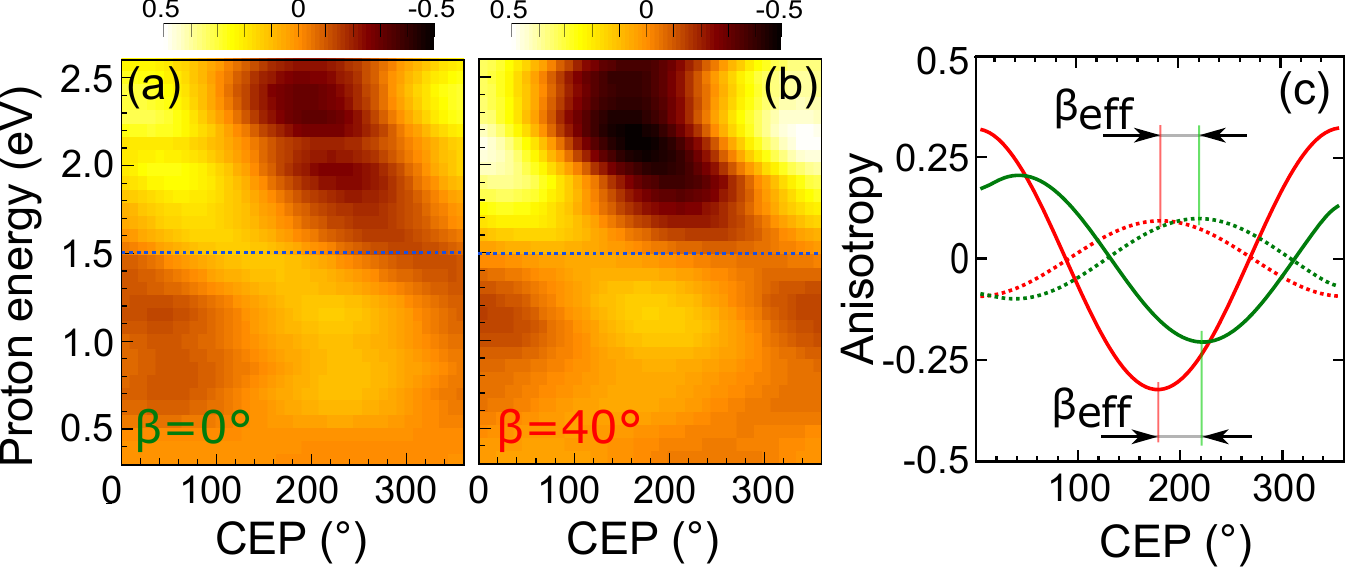}
    \caption{(a,b) Lab frame anisotropy $\labfaniso(\cep,E_k)$ for proton ejection under $\beta=0(\pm10)^\circ$ (a) and $\beta=40(\pm10)^\circ$ (b). (c) $\labfaniso$ over \CEP{} for $\beta=0^\circ$ (green) and $\beta=40^\circ$ (red), plotted for $E_k>1.5$~eV (full lines) and $0.6 < E_k < 1.5$~eV (dashed lines). $\beta_\textrm{eff} \approx 36^\circ$ is indicated.}
  \label{fig2}
\end{figure}


Fig.~\ref{fig1}(c) shows that for $E_k>0.7$\,eV in our experiment   $\molfaniso<0$ for $0^\circ <\imphase<180^\circ$, and $\molfaniso>0$ for $180^\circ <\imphase<360^\circ$
This means that the proton is much more likely ejected against the electron emission direction, i.e., $\vec{p}_e \cdot \protonmom <0$, than with the electron, independent of the alignment of the molecule in the lab frame. 
That holds for all values of the CEP, see \SM{}. 
For $0.4<E_k<0.7$\,eV the situation is inverted and the proton is more likely ejected into the same hemisphere as the electron. 
Our experimental anisotropy for a few-cycle pulse is only qualitatively similar to that of Ref.~\cite{Wu2013} for a multi-cycle pulse. 
On a quantitative level we measure much higher values of $|\molfaniso|$ (by about a factor of 6) due to the larger intensity and the shorter pulse duration \cite{Alnaser2004_H2_coulomb_explosion, Staudte2007, Hanus2019,  Hanus2020} in our experiment,  and almost no dependence of $\molfaniso$ on $E_k$ for $E_k>0.7$\,eV. 
The latter we attribute to the much broader spectrum of our laser pulse, which enables transitions between the 1s$\sigma_g$ and 2p$\sigma_u$ states over a much broader range of internuclear distances $R$ [cf. Fig.~\ref{fig1}(f)] resulting in averaging over a broader range of $E_k$.


%
%
%

The result in Fig.~\ref{fig1}(c), that the ejection direction of a proton with a given $E_k$ is determined to a large degree only by the direction of electron emission, implies a strong correlation between the outgoing electron and the ejected proton. This correlation is mediated by the laser-sub-cycle intramolecular dynamics of the bound electron. In turn, it may provide experimental access to this dynamics, see \SM{} for details. 
Within the abovementioned semi-classical picture of pathway interferences the correlation can be explained as follows.
For a given $E_k$ and photon energy $\omega$, the delay $\Delta t =  t_{n\omega} - t_i$ between ionization time $t_i$ and the times of transition(s) between the two states at $t_{n\omega}$ [cf. Fig.~\ref{fig1}(f)] is constant for a specific $n$  \cite{Ji2019}. 
In turn, also the quantum phase difference $\Delta \varphi$ between two pathways with $n$ and $n+1$ is constant.
Therefore, $\Delta \varphi$ will invert its sign for $\imphase \rightarrow\imphase+180^\circ$, which entails a change of the anisotropy $\molfaniso \rightarrow -\molfaniso$.
Thus, for a given photon energy $\omega$, the dependence of the proton anisotropy $\molfaniso$ on $\imphase$ is an \textit{intrinsic} property of H$_2^+$ determined by its potential energy structure.

\begin{figure*}[tb]
 \begin{center}
 \includegraphics[width=\textwidth]{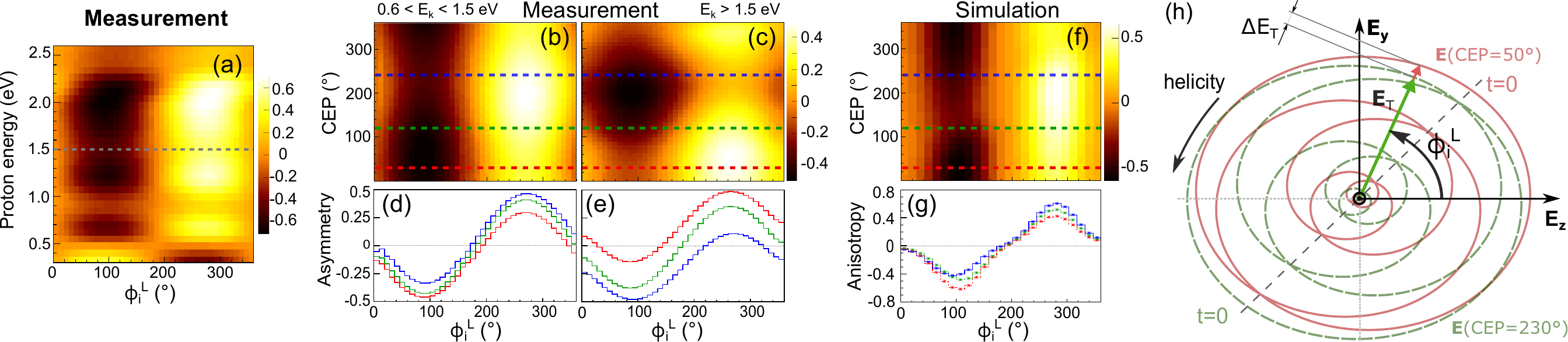}
  \end{center}
   \caption{
   (a) Measured proton ejection anisotropy $\labfaniso^{\beta=0}$ as defined in the text, over the laser phase at ionization time, $\ilphase$, and proton kinetic energy $E_k$, integrated over the \CEP{}. 
   (b,c) $\labfaniso^{\beta=0}$ over $\ilphase$ and $\cep$ for $0.6<E_k<1.5$~eV (b) and $E_k>1.5$~eV (c). 
   (d) Lineouts from (b) for three values of the \CEP{}: 30$^\circ$, 120$^\circ$, 240$^\circ$ (from topmost to bottommost line). 
   (e) Same as (d) but lineouts from (c). 
   (f) Simulated $\labfaniso^{\beta=0}$ for $E_k=1$~eV corresponding to (b). 
   (g) Lineouts from (f), description as in (d). 
   (h) Evolution of the laser electric field vector $\vec{E}(t)$ in the lab frame for $\cep=50^\circ$ (red, full) and $\cep=230^\circ$ (green, dashed). The laser field vectors at transition time, $\vec{E}_T = \vec{E}(t_T)$, are indicated for both field forms together with their difference, $\Delta E_T$.}
   \label{fig3}
\end{figure*}

Even though the high values of the intrinsic anisotropy $\molfaniso(\imphase, E_k)$ in Fig.~\ref{fig1}(c) suggest a promising handle for directional control over the bond-breaking process using the phase $\imphase$, the anisotropy only appears in the molecular frame and cannot be exploited for lab frame control: The anisotropy changes sign for $\imphase \rightarrow \imphase + 180^\circ$. 
Thus, in a randomly oriented ensemble of symmetric molecules, for  every proton ejection angle $\beta$ there exists   an angle $\beta' = \beta + 180^\circ$ for which the same number of protons are ejected into the opposite direction. As a result, the anisotropy in the molecular frame described in Ref.~\cite{Wu2013} and depicted in Fig.~\ref{fig1}(c)  vanishes for integration over $\imphase$. 
Moreover, the quantity $\imphase$ is accessible only in electron-ion coincidence experiments and can only be measured but not controlled in an experiment. 
Only a lab frame quantity that can be adjusted using an experimental knob, such as the \gls{cep}, can be used for bond-breaking control in the lab frame.





Still, the fact that the proton ejection direction is to a large degree determined by the ionization step, actually questions the role of the CEP-determined laser field's shape during the laser-molecule interaction. 
To clarify this question, we  turn to investigating the connection between the molecular frame anisotropy $\molfaniso$, shown in Fig.~\ref{fig1}(c), and the CEP-dependent lab frame anisotropy, shown in Figs.~\ref{fig1}(d,e). For this, we first need to connect the molecular frame with the lab frame. 
To see this connection, we calculate the laser electric field along the molecular axis:  $E_\textrm{mol}(t,\cep,\beta) = \hat{E}_z f(t) \sqrt{\cos^2{\beta}+\varepsilon^2 \sin^2{\beta}} \cos({\omega t+\cep-\beta_\textrm{eff}})$ with $\beta_\textrm{eff}=\tan^{-1}({\varepsilon \tan{\beta}})$. 
This relation tells us that for a given ionization time $t_i$, the alignment angle $\beta$ and the \gls{cep} lead to equivalent changes of the molecular frame ionization phase $\imphase = \omega t_i+\cep-\beta_\textrm{eff}$.

As $\beta$ and $\cep$ are both lab frame quantities, they connect the molecular frame phase $\imphase$ with the lab frame. To establish this connection, we turn to the normalized lab frame proton anisotropy, $\labfaniso(\cep,E_k,\beta)$, which we define in accordance with $\Delta \momdis$ in Figs.~\ref{fig1}(d,e). After transformation from the cartesian $(\protonmomz, \protonmomy)$ to the polar coordinates $(E_k, \beta)$ we obtain 
$\labfaniso(\cep,E_k,\beta) = \frac{\Delta \momdis(\cep,E_k,\beta)-\Delta \momdis(\cep,E_k,\beta+180^\circ)}{\momdis(\cep,E_k,\beta)+\momdis(\cep,E_k,\beta+180^\circ)}$.
The such defined anisotropy is shown in Figs.~\ref{fig2}(a,b) for $\beta=0$ and $\beta=40^\circ$. 
Fig.~\ref{fig2}(c) shows the anisotropy  obtained by integrating over the low and high $E_k$ regions for which $\labfaniso(\cep,E_k,\beta)$ shows opposite \CEP{}-dependence.
In both $E_k$ regions, a rotation of the molecule by $\beta = 40^\circ$ results with $\varepsilon=0.85$ in a \gls{cep}-shift of 
$\beta_\textrm{eff}=\tan^{-1}({0.85 \tan(40^\circ)}) \approx 36^\circ$, 
proving the equivalence of $\beta$ and $\cep$.


To connect the molecular and lab frame anisotropies $\molfaniso$ and $\labfaniso$ we need to use a lab frame counterpart of $\imphase$,  the phase defining the direction of $\vec{E}(t_i)$ in the molecular frame. From Fig.~\ref{fig1}(a) we find the angle of $\vec{E}(t_i)$ in the lab frame as $\ilphase=\imphase+\beta=\omega t_i + \cep$. Because of the equivalence of $\beta$ and \CEP{} established in Fig.~\ref{fig2}, in order to make explicit the influence of the \CEP{} and not to smear out its action, we need to fix the value of $\beta$. 
If $\beta$ is random, the influence of the \CEP{} in the proton anisotropy gets suppressed. 
Without loosing generality, we set $\beta=0^\circ$, for which the connection between $\ilphase$ and $\imphase$ becomes particularly simple, namely $\ilphase = \imphase|_{\beta=0}$.
With that, we can straightforwardly adopt the definition of $\molfaniso$ and obtain $\labfaniso^{\beta=0}(\ilphase,\cep,E_k) = \frac{N(\ilphase,\cep,E_k,\beta=0)-N(\ilphase,\cep,E_k,\beta=180^\circ)}{N(\ilphase,\cep,E_k,\beta=0) + N(\ilphase,\cep,E_k,\beta=180^\circ)}$, where we have made explicit the dependence of $\labfaniso^{\beta=0}$ on the \CEP{}.
The anisotropy  $\labfaniso^{\beta=0}$ is an equivalent but generalized form of the anisotropy  $\molfaniso$ defined above. 
We can see this equivalence if we replace $\imphase \rightarrow \ilphase-\beta$ in $\molfaniso(\imphase,E_k)$. For $\beta = 0$ this is exactly the definition of $\labfaniso^{\beta=0}$, but with the additional dependence on CEP.

We plot the measured $\labfaniso^{\beta=0}(\ilphase,\cep,E_k)$, integrated over \CEP{}, in  Fig.~\ref{fig3}(a). It closely resembles $\molfaniso(\imphase,E_k)$ that is integrated over all angles $\beta$, shown in Fig.~\ref{fig1}(c).
Its lab frame counterpart for $\beta=0$, $\labfaniso^{\beta=0}(\ilphase,\cep,E_k)$, finally enables us to investigate the separate actions of the ionization phase $\ilphase$ and the $\cep$ as well as their connection. 
To this end, we plot $\labfaniso^{\beta=0}(\ilphase,\cep,E_k)$ in Figs.~\ref{fig3}(b,c), separated into high and low proton energy ranges that show opposite CEP-dependence.
The two energy ranges clearly visible in Figs.~\ref{fig2}(a,b) correspond to the overlap of the $1\omega$ and $2\omega$, and the $2\omega$ and $3\omega$ pathways, respectively. 
The larger peak value of $\labfaniso$ for $\beta=40^\circ$ is attributed to a more favorable population ratio of the interfering dissociation pathways for this angle,  due to the $\beta$-dependence of the intensity along the molecular axis $|E_\textrm{mol}|^2$. 
The distributions in Fig.~\ref{fig3}(b,c) show that in both energy regions $\ilphase$ has a dominant influence on the proton ejection direction. However, the \gls{cep} modulates the value of $\labfaniso^{\beta=0}$, evident from its variation along the \gls{cep}-axis and the corresponding cuts along $\ilphase$ for selected values of the \gls{cep} in Fig.~\ref{fig3}(d,e).
In these lineouts the \gls{cep}-induced offsets in $\labfaniso^{\beta=0}(\ilphase)$ are clearly visible. 

Thus, the action of the \gls{cep} for determining the proton anisotropy is a modulation of the intrinsic anisotropy due to the ionization phase $\ilphase$. 
To elucidate the mechanism behind this \gls{cep}-modulation we developed a simple semi-classical model that calculates the anisotropy-determining phase difference $\Delta \varphi$
acquired by a vibrational wavepacket along the different dissociating pathways.  
For a qualitative assessment of the physics underlying the measured $\labfaniso^{\beta=0}(\ilphase,\cep,E_k)$ we  restricted our model to the low proton energy region where only the $1\omega$ and $2\omega$ pathways interfere, see \SM{} for details. 
The anisotropy map $\labfaniso^{\beta=0}(\ilphase,\cep,E_k)$ predicted by this model, displayed in Fig.~\ref{fig3}(f), as well as the lineouts for selected values of the \gls{cep}, shown in Fig.~\ref{fig3}(g), resemble the measured quantities in Figs.~\ref{fig3}(b) and (d) to a remarkable degree.

This good agreement, despite the dedicated qualitative character of the model, is due to the correct incorporation of the key mechanism underlying the influence of the \gls{cep} on the anisotropy, which is the pronounced variation of the electric field strength with \gls{cep} at the times when the \ssg{} and \psu{} states  are coupled by $N$-photon transitions, see \SM{} for details.
%
In short, the transition probability $P$ between the two electronic states is proportional to the field strength in the molecular axis at transition time $t_T$, according to $P \propto |{E}_\textrm{mol}(t_T)|^{2N}$. 
%
%
Thus, the \CEP{}-variation of the transition probability $P$, and therewith that of the anisotropy-determining phase difference $\Delta 
\varphi$, is determined by the \CEP{}-modulation of $E_\textrm{mol}(t_T)$, cf. the visualization of this field-variation with \gls{cep} in Fig.~\ref{fig3}(h). 
%
Since the \CEP{}-variation of $E_\textrm{mol}(t_T)$ decreases with the number of cycles in the pulse, the modulation of the proton anisotropy with the \CEP{}, visible in Figs.~\ref{fig3}(d,e,g), vanishes for a multi-cycle pulse. As a result, for a multi-cycle pulse only the  dependence of the anisotropy on the ionization phase $\imphase$, which can only be detected in a coincidence experiment, remains. 
%


In conclusion, we investigated the phase-sensitivity of bond-breaking in dissociative laser-ionization of H$_2$. We establish a unified picture that relates the well-known \CEP{}-dependence of the proton anisotropy in the lab frame measured with few-cycle pulses \cite{Kling2006, Kremer2009, Fischer2010, Znakovskaya2012, Xu2013,  Fischer2013, Kling2013a, Li2016} to an intrinsic proton anisotropy that depends on the laser phase at ionization time. 
Our work shows that the former anisotropy arises due to a \CEP{}-modulation of the latter via the \CEP{}-dependence of the transition amplitudes between interfering nuclear pathways. 
Our experiments also reveal a remarkably strong correlation between the outgoing electron and the ejected proton that is directly connected to the intramolecular electron dynamics during dissociation. 
We predict that this correlation opens up a new approach to access the laser-sub-cycle electronic dynamics during molecular bond-breaking, see \SM{} for a detailed reasoning.
%




\acknowledgments
This work was supported by the Austrian Science Fund (FWF), Grants No. 
P28475-N27  
and P30465-N27,   
by a MEXT (Japanese Ministry of Education, Culture, Sports, Science and Technology) Grant-in-Aid for Specially Promoted Research (No. JP15H05696), 
and by the National Natural Science Foundation of China (NSFC) (Grant No. 11574205).
We gratefully acknowledge discussions with Jian~Wu, Uwe~Thumm and Andr\'{e}~Staudte.


%

\end{document}